\documentstyle[12pt,prc,aps,psfig]{revtex}
\tighten
\begin{document}
\draft

\title{Predicting the single-proton/neutron potentials \\
in asymmetric nuclear matter}
\author{F. Sammarruca, W. Barredo, and P. Krastev}
\address{Physics Department, University of Idaho, Moscow, ID 83844, U.S.A}
\date{\today}
\maketitle
\begin{abstract}
We discuss the one-body potentials for protons and neutrons obtained from 
 Dirac-Brueckner-Hartree-Fock calculations of neutron-rich matter, in particular      
 their dependence upon the degree of proton/neutron asymmetry. The closely 
 related symmetry 
 potential is compared with empirical information from the isovector component
 of the nuclear optical potential. 
\end{abstract}
\narrowtext

\section{Introduction} 

Lately we have been concerned with           
probing the behavior of the isospin-asymmetric
equation of state (EOS) \cite{AS02}. With our recent work on neutron radii and neutron skins
\cite{AS203}, we have explored applications of the EOS at densities typical for normal nuclei.

 It is also important to look into                        
systems that are likely to constrain the behavior of the EOS at higher
densities, where the largest model dependence is observed.               
Supernova explosions and neutron star formation/stability are phenomena where the nuclear EOS
plays a crucial role.  The symmetry energy determines the proton fraction in neutron 
stars in $\beta$ equilibrium, and, in turn, the cooling rate and neutrino emission.
 Models of prompt supernova explosion and systematic analyses of
neutron star masses provide often conflicting information on the ``softness" of the 
EOS and its incompressibility at equilibrium.         

On the other hand, 
 collisions of neutron-rich nuclei, which are the purpose of 
the Rare Isotope Accelerator (RIA), provide a unique opportunity to obtain terrestrial data 
suitable for constraining the properties of dense and highly asymmetric matter.
Such reactions are capable of producing extended regions
of space/time where both the total nucleon density and the neutron/proton 
asymmetry are large.                                                               
Isospin-dependent Boltzmann-Uehling-Uhlenbeck (BUU) transport models \cite{BUU} 
include isospin-sensitive collision dynamics through the elementary $pp$, $nn$, and $np$
cross sections
and the mean field. The latter is a crucial isospin-dependent mechanism,
and is the focal point of this paper. 
The contribution to the mean field from the neutron/proton asymmetry                 
can be
 measured through isospin-sensitive observables \cite{BAL}. 
 In summary, this is a timely and exciting topic, which is  
 stimulating new effort, on both the experimental and the theoretical sides.  

  At this time, it is fair to say that the model dependence of the isospin
asymmetric EOS is rather large. In fact, even the qualitative behavior  of some 
predictions is            
 controversial, as is the case, for instance, with 
 the density dependence of the symmetry energy, upon which isospin diffusion in heavy-ion
 collisions is found to depend sensitively \cite{u01}.
 Thus any additional constraint is 
 desirable and should be fully explored.  
 As discussed in Ref.~\cite{BAL04}, nucleon-nucleus optical
potential information can be exploited to constrain the strength and the energy dependence of the 
single-neutron/proton potentials                                                       
in asymmetric nucler matter.                                  
 The basic idea is that, even though infinite nuclear matter
 is an idealized system, the single-nucleon potentials should bear a clear
signature of the optical potential in the interior of the nucleus.

In this paper we will discuss the predictions for the single-neutron/proton
potentials and the closely related symmetry potential as obtained from  
our Dirac-Brueckner-Hartree-Fock
calculations of asymmetric matter. We will compare with other predictions from the literature as well as                     
empirical optical potential
information.             
We will point out the large model dependence of predictions for those observables that depend 
sensitively on the difference between neutron and proton properties in asymmetric
matter.         
Additional experimental constraints are therefore important. Moreover,     
microscopic, parameter-free approaches are the best way to gain deeper
insight into the isospin-dependent properties of nuclear matter.             

\section{The single-nucleon potentials} 
\subsection{Momentum dependence} 

Unless otherwise specified, we use  
 the Bonn-B potential \cite{Mac89}                                    
and the relativistic Brueckner-Hartree-Fock (DBHF) model outlined in Ref.~\cite{AS02}.

We begin by examining the momentum dependence of 
$U_{n/p}$, the single neutron/proton potential in neutron-rich matter. 
In Fig.~1, we show 
$U_{n/p}$ as a 
function of the momentum and for different values of the asymmetry parameter,
$\alpha=(\rho_n - \rho _p)/(\rho_n + \rho_p)$, with $\rho_n$ and $\rho_p$ the 
neutron and proton densities. 
The total nucleon density considered in the figure is equal to 0.185 fm$^{-3}$ and
corresponds to a Fermi momentum of 1.4 fm$^{-1}$, which is very close to our predicted
saturation density.

\begin{figure}
\begin{center}
\vspace*{0.5cm}
\hspace*{-0.5cm}
\psfig{figure=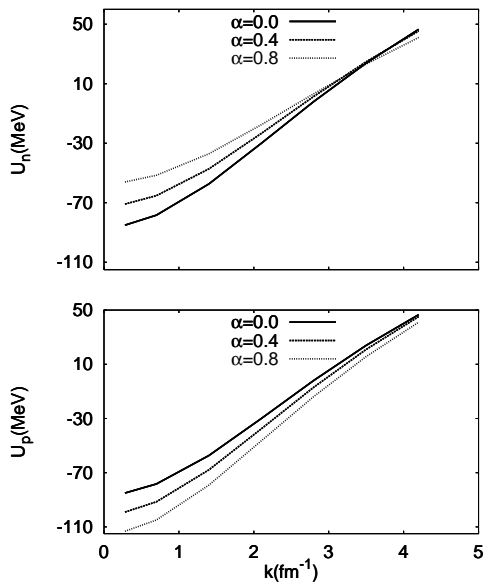,height=8.0cm}
\vspace*{0.5cm}
\caption{ The single-neutron (upper panel) and single-proton (lower panel)
potential as a function of the nucleon momentum for three different values
of the asymmetry parameter. The average Fermi momentum is 1.4 fm$^{-1}$. 
} 
\label{one}
\end{center}
\end{figure}

 For increasing values of 
$\alpha$, the proton potential becomes increasingly attractive while the opposite
tendency is observed in $U_n$. This reflects the fact that the proton-neutron interaction, the 
one predominantly felt by the single proton as the proton density is depleted, is more
attractive than the one between identical nucleons. 
Also, as it appears reasonable, the dependence on $\alpha$ becomes weaker at larger momenta.

\begin{figure}
\begin{center}
\vspace*{0.5cm}
\hspace*{-0.5cm}
\psfig{figure=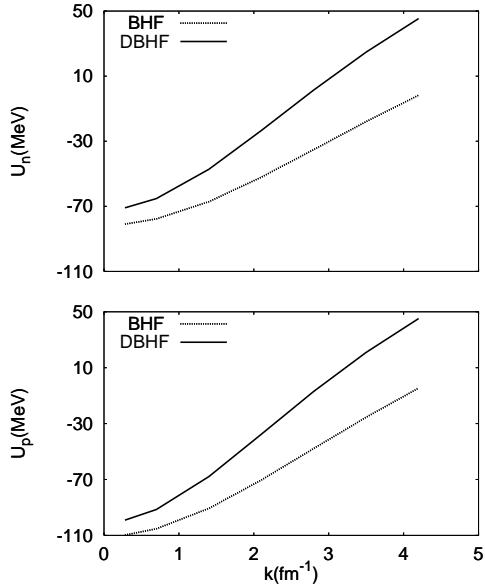,height=8.0cm}
\vspace*{0.5cm}
\caption{ Comparison between DBHF and BHF predictions of the 
single-neutron (upper panel) and single-proton (lower panel)
potential. The value of the asymmetry parameter is fixed to 0.4 and 
the average Fermi momentum is 1.4 fm$^{-1}$. 
} 
\label{two}
\end{center}
\end{figure}
In Figure 2 we show the DBHF results in comparison with those from 
(non-relativistic) conventional Brueckner-Hartree-Fock (BHF) calculations.             
 We make the comparison to
show the considerable difference between the two sets of results as well as to check that our BHF
predictions are 
in qualitative agreement with other studies based on the conventional Brueckner
G-matrix approach. An older work based on that approach can be found, for instance, in Ref.~\cite{BL91}, 
where separable representations of the nucleon-nucleon interaction are adopted. More
recent calculations have been reported 
in Ref.~\cite{Tub04}, where the CD-Bonn potential \cite{CD} 
is used in conjunction with the BHF approximation. 

The role of the momentum dependence of the symmetry potential in heavy-ion collisions was recently
examined \cite{DAS04} and found to be important. Symmetry potentials with and without 
momentum dependence and yielding similar predictions for the symmetry energy can lead to significantly
different predictions of collision observables \cite{DAS04}.

\subsection{Asymmetry dependence and the symmetry potential} 

Regarding                                                                      
$U_{n/p}$ as functions of                                    
the asymmetry parameter $\alpha$, one can easily
verify that the following approximate relation applies                 
\begin{equation}
U_{n/p}(k,k_F,\alpha) \approx U_{n/p}(k,k_F,\alpha=0) \pm U_{sym}(k,k_F)\alpha 
\end{equation}
with the $\pm$ referring to neutron/proton, respectively.                   
Figure 3 displays the left-hand side of Eq.~(1) for fixed density and nucleon momentum and
clearly reveals the linear 
behaviour of                                   
$U_{n/p}$ as a function of $\alpha$.                       
\begin{figure}
\begin{center}
\vspace*{0.5cm}
\hspace*{-0.5cm}
\psfig{figure=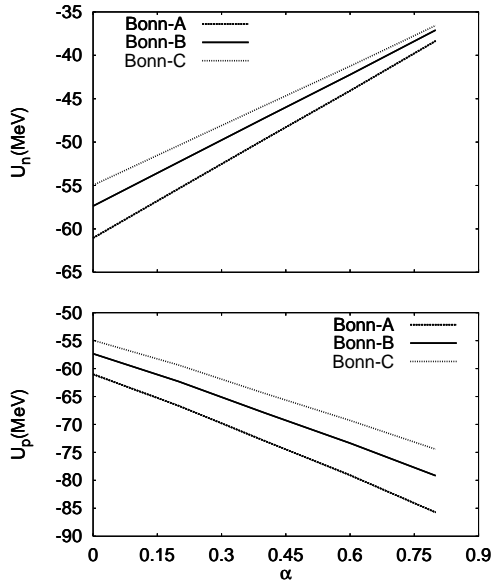,height=8.0cm}
\vspace*{0.5cm}
\caption{ The single-neutron (upper panel) and single-proton (lower panel)
potential as a function of the asymmetry parameter for fixed average density
($k_F$= 1.4 fm$^{-1}$) and nucleon momentum  
($k$= $k_F$).                                 
} 
\label{three}
\end{center}
\end{figure}

Although the main focus of Fig.~3 is the $\alpha$ dependence, 
predictions are displayed for the Bonn A, B, and C potentials \cite{Mac89}.
These three models differ mainly in the strength of the tensor force, which
is mostly carried by partial waves with isospin equal to 0 and thus should fade away
in the single-neutron potential
as the neutron fraction increases. Reduced differences among the three models are in fact observed              
in $U_n$ at the larger values of $\alpha$.

Already several decades ago, it was pointed out that the real part of the nuclear 
optical potential depends on the asymmetry parameter as in Eq.~(1) \cite{Lane}. 
Thus,                                                                   
the quantity 
\begin{equation}
\frac{U_{n} + U_p}{2} = U_0 ,               
\end{equation}
which is obviously the single-nucleon potential in absence of asymmetry,
should be a reasonable approximation to the isoscalar part of the optical 
potential. The momentum dependence of $U_0$ (which is shown in Fig.~1 as the 
$\alpha$=0 curve), is important for extracting information about the symmetric matter EOS
and is reasonably agreed upon
 \cite{u01,u02,u03,u04,u05,u06,u07,u08,u09}.

On the other hand, 
\begin{equation}
\frac{U_{n} - U_p}{2\alpha} = U_{sym}             
\end{equation}
should be comparable with the Lane potential \cite{Lane}, or the isovector 
part of the nuclear optical potential \cite{Lane}. (Notice that in the two equations
above the dependence upon density, momentum, and asymmetry has been suppressed for
simplicity.) 
We have calculated $U_{sym}$ close to nuclear matter density and as a function of
the momentum, or rather the corresponding kinetic energy. 
The predictions obtained with Bonn A, B, and C are shown in Fig.~4.
They are compared with the phenomenological expression \cite{Lane}
\begin{equation}
U_{Lane}      = a -b T                            
\end{equation}
where $T$ is the kinetic energy, $a \approx 22-34 MeV$, $b\approx 0.1-0.2 MeV$.
\begin{figure}
\begin{center}
\vspace*{4cm}
\hspace*{-0.5cm}
\psfig{figure=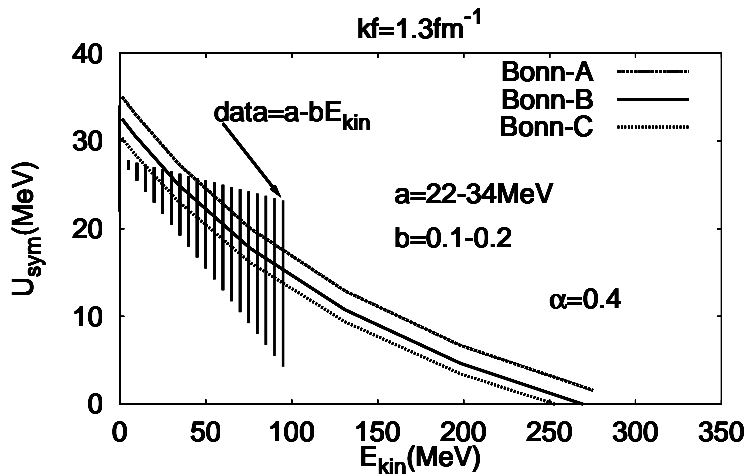,height=6.5cm}
\vspace*{-2.5cm}
\caption{ The symmetry potential as a function of the nucleon kinetic energy at     
nuclear matter density. The predictions obtained with Bonn A, B, and C are
compared with empirical information from nuclear optical potential data (shaded area). See text
for details. 
} 
\label{four}
\end{center}
\end{figure}

The strength of the predicted symmetry potential decreases
with energy, a behavior which is consistent with the empirical information.   
The same comparison is done in 
  Ref.~\cite{BAL04} starting from a                             
  phenomenological formalism for the single-nucleon potential \cite{Bomb01,Rizzo}. 
There, it is shown that it is 
 possible to choose two sets of parameters which lead to similar values of the 
 symmetry energy but exactly opposite tendencies in the energy dependence of the symmetry
 potential as well as 
 opposite sign of the proton-neutron mass splitting.
 As a consequence of that, these two sets of parameters lead to very different
 predictions for observables in heavy-ion collisions induced by neutron-rich nuclei
 \cite{Rizzo}.

Our effective masses for proton and neutron are shown in Fig.~5 as a function
of $\alpha$ and at saturation density.                                          
 The predicted effective mass of the neutron being larger than the 
proton's is a trend shared with 
microscopic non-relativistic
calculations \cite{BL91}. In the non-relativistic case, one can show from very elementary arguments
based on the curvature of the single-particle potential that a more attractive 
potential, as the one of the proton, leads to a smaller effective mass.
In our DBHF effective-mass approximation, we 
assume momentum-independent nucleon self-energies, $U_S$ and $U_V$, with a vanishing
spacial component of the vector part. In such limit, following similar calculations
of symmetric matter \cite{BM}, the one-body potential is written as \cite{AS02}
\begin{equation}
U_i(p) = \frac{m^*_i}{E^*_i}U_{S,i} + U_{V,i}       
\end{equation}
where $E^*_i=\sqrt{(m^*_i)^2 + p^2}$, 
$m^*_i = m_i + U_{S,i}$, and $i=n$ or $p$ for neutrons or protons, respectively.
Defining for convenience
$U_{0,i} = U_{S,i} + U_{V,i}$,     
the expression above becomes a two-parameter formula which requires the fitting of two
constants, just like in the non-relativistic case. 
Now, since the single-proton potential is more attractive (see Fig.~1), and both 
the neutron and proton potentials tend to the same limit at high momenta, it is easy
to see from Eq.~(5), or rather its derivative, that the proton effective mass obtained
in this way must 
be smaller than the neutron's. 

\begin{figure}
\begin{center}
\vspace*{4.0cm}
\hspace*{-0.5cm}
\psfig{figure=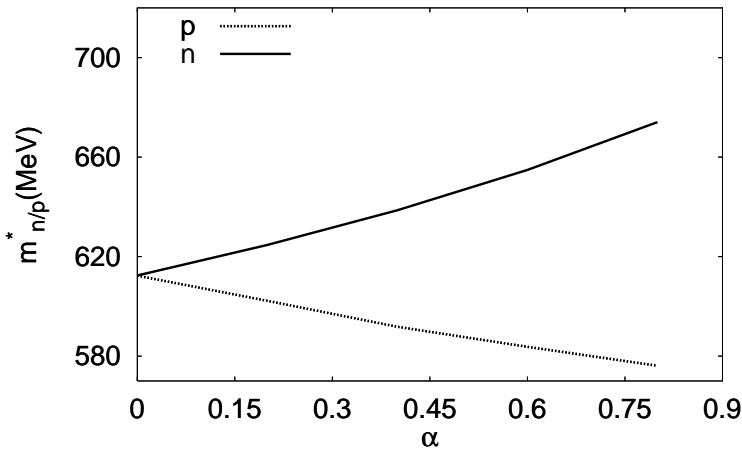,height=6.5cm}
\vspace*{-2.5cm}
\caption{ The proton and neutron effective mass as a function of the asymmetry
parameter and for fixed average density 
($k_F$= 1.4 fm$^{-1}$).                                 
} 
\label{five}
\end{center}
\end{figure}
One can encounter in the literature the statement that relativistic mean field (RMF) models
predict a switch in sign of the neutron-proton effective mass splitting whenever the 
scalar isovector $\delta$ meson is included \cite{Rizzo}. We suggest                          
that the reason for this mechanism is in the first-order nature of RMF calculations.
The first-order contribution of the $\delta$ meson is known to be 
repulsive in $^{3}S_{1}$ (the most important partial wave for generating nuclear binding).
Therefore, in a context where the proton density increases (thus making the $np$ interaction
the dominant contribution to the single-proton potential), a first-order calculation would in fact
generate additional {\it repulsion} in the single-proton potential. Hence the larger 
proton effective mass. On the other hand, in more microscopic approaches, such as the one
we use, the nucleon potential is iterated to all orders via the in-medium scattering equation.
In particular, most of the attraction in the $np$ interaction is generated in second order,
through the two-pion exchange. (The $\delta$ meson is included, of course, but mainly for the 
sake of accurate reproduction of NN phase shifts.) As we already observed in Section IIA,
increasing the $np$ contribution to the single-proton potential results in increased attraction.

\begin{figure}
\begin{center}
\vspace*{0.5cm}
\hspace*{-0.5cm}
\psfig{figure=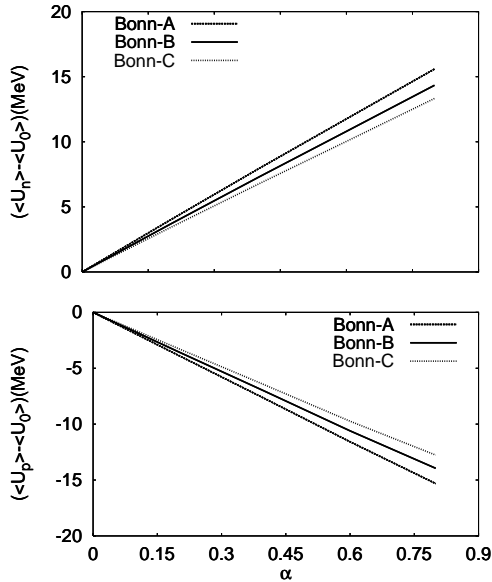,height=8.0cm}
\vspace*{0.5cm}
\caption{                                                                    
Contribution from the asymmetry to the average potential energy per neutron
(upper panel) and proton (lower panel). 
Average density as in the previous figures.
}
\label{six}
\end{center}
\end{figure}
 Some comments are in place concerning other DBHF 
 calculations of asymmetric matter where the Dirac 
 neutron effective mass is reported to be larger 
 than the proton's \cite{plb,DFF}. In Ref.~\cite{plb}, the authors argue
 that the Dirac mass, $m^* = m + U_S$, should not be
 compared with the effective mass which can be extracted,
 for instance, from analyses based on non-relativistic
 optical models. Instead, an effective mass based on the
 energy dependence of the Shroedinger equivalent potential 
 should be calculated, in which case one finds that             
$m^*_n >m^*_p$.
 The arguments found in Ref.~\cite{plb} are well known and 
 were already advanced in Ref.~\cite{fw}, where as many as six
 different definitions for the effective mass are introduced.
 Again, one must keep in mind the point we already made above. If the 
 nucleon self-energy is written in terms of a scalar potential and
 {\it only} the time-like component of a vector potential, and both are
 taken to be momentum independent, it is then easy to see that
 the expansion of the single-particle energy is consistent to 
 leading order with the non-relativistic single-particle energy.
 Thus it is reasonable, and in fact to be expected, that                            
 our Dirac masses would be qualitatively
 consistent with those from non-relativistic predictions, such as
 BHF calculations. 
 In summary, the effective mass is just part of a convenient
 parametrization of the single-particle potential. Clearly, how
 many terms are retained in the nucleon self-energy and how their
 momentum dependence is handled will impact the parametrization.
 Ultimately, physical observables depending on the neutron/proton
 mean field  must be correctly described, irrespective of the 
 chosen parametrization.

One conclusion that can be made from all of the above is that 
 the parameters of the single-nucleon potential in 
asymmetric matter 
appear to be weakly correlated to observables
such as the energy per particle or the symmetry energy, where proton and neutron
contributions are averaged together. 
Constraints from ``differential" or relative observables, namely those  specifically sensitive to the difference
between proton and neutron properties, are thus very much needed.

Before closing, we also show for completeness the average potential energy per neutron/proton,
where the momentum dependence has been integrated out. This is the              
	proton/neutron potential 
energy contribution to the total energy per particle which then appears in the EOS.
Actually, what we show in Fig.~5  are the 
average potential energies from which 
the part coming from the symmetric EOS has been subtracted out, that is, just 
  the contribution from the 
asymmetry to the interaction potential energy, 
\begin{equation}
<\Delta U_{n/p}>(\rho,\alpha) = <U_{n/p}>(\rho, \alpha) - 
 <U (\rho, \alpha=0)>.       
\end{equation}

Clearly, the contribution from the asymmetry, in both the momentum-dependent and
the momentum-averaged potentials, turns
out to be large and positive for
neutrons, large and negative for protons. This component of the mean field will then
be effective in 
separating the collision dynamics for neutrons and protons by 
making more neutrons unbound than protons (or, by making the neutrons more energetic, if
already unbound). This effect can be discerned through observables such as the
neutron/proton differential flow in heavy-ion collisions \cite{BAL}.

\section{Conclusions} 
We have focussed on some of the properties of neutrons and protons in neutron-rich matter.
This is a topic of contemporary interest. Its relevance extends from the
dynamics of colliding nuclei to nuclear astrophysics.

Different models may be in fair agreement with respect to averaged properties of 
the EOS, and yet produce very different predictions of properties such as the 
symmetry potential and the closely related single-nucleon potentials and effective
masses. 
Clearly, more stringent constraints are needed for the 
isospin-dependent properties of the EOS.

Very good transport model calculations are available from the literature                  
\cite{BUU,BAL}. However, considerable
amount 
of phenomenology is often involved in the input of these models (for instance, the mean field is based on some  phenomenological interaction \cite{DAS03,DAS04}
and/or the elementary cross sections are obtained from empirical data). 
We calculate all of the above ingredients {\it microscopically} and 
internally consistent
 with respect to the two-body force.     
 We are presently studying the dependence on density {\it and} asymmetry of the in-medium 
 isospin-dependent nucleon-nucleon cross sections with the purpose of obtaining a
 convenient parametrization as a function of energy, density, and asymmetry.
 Our microscopic information (both elementary cross sections and mean field) can be a valuable 
input for transport model calculations of                                              
heavy-ion dynamical observables.
  This combined effort will complement  
 new data to 
be taken at RIA and eventually shed light on the less known aspects of the nuclear       
equation of state. \\ \\ 
\begin{center}
{\bf ACKNOWLEDGMENTS}
\end{center}
The authors acknowledge
financial support from the U.S. Department of Energy under grant number DE-FG02-03ER41270.

\end{document}